\documentclass[usenatbib]{mnras}
 
\usepackage[utf8]{inputenc}
\usepackage[T1]{fontenc}
\usepackage{ae,aecompl}

\usepackage{ulem}
\usepackage{amsmath}
\usepackage{eqnarray}
\usepackage{graphicx}
\usepackage{tabularx}
\usepackage{placeins}
\usepackage{siunitx}
\usepackage{titlesec}
\usepackage{epstopdf}
\usepackage{flafter}
\usepackage{multirow}
\usepackage{appendix}

\newcommand{\HII}{\mbox{H\,\textsc{ii}}}%
\usepackage{color}

\newcommand{\mum}{$\mu$m}

\graphicspath{ {./pca_figures/} }

\title[A PCA of PAH emission]{A Principal Component Analysis of polycyclic aromatic hydrocarbon emission in NGC~2023}

\author[A. Sidhu et al.]{Ameek Sidhu$^{1,2}$\thanks{E-mail: asidhu92@uwo.ca}, Els Peeters$^{1,2,3}$, Jan Cami$^{1,2,3}$ and Collin Knight$^{1}$\\
$^{1}$Department of Physics \& Astronomy, University of Western Ontario, London, ON, N6A 3K7, Canada\\
$^{2}$Institute for Earth and Space Exploration, University of Western Ontario, London, ON, N6A 3K7, Canada\\
$^{3}$SETI Institute, 189 Bernardo Avenue, Suite 100, Mountain View, CA 94043, USA}

\begin{document}

\date{}

\maketitle

\begin{abstract}
We use the measured fluxes of polycyclic aromatic hydrocarbon (PAH) emission features at 6.2, 7.7, 8.6, 11.0 and 11.2 $\mu$m in the reflection nebula NGC~2023 to carry out a principal component analysis (PCA) as a means to study previously reported variations in the PAH emission. We find that almost all of the variations (99\%) can be explained with just two parameters -- the first two  principal components (PCs). We explore the characteristics of these PCs and show that the first PC ($PC_{1}$), which is the primary driver of the variation, represents the amount of emission of a mixture of PAHs with ionized species dominating over neutral species. The second PC ($PC_{2}$) traces variations in the ionization state of the PAHs across the nebula. Correlations of the PCs with various PAH ratios show that the 6.2 and 7.7 $\mu$m bands behave differently than the 8.6 and 11.0 $\mu$m bands, thereby forming two distinct groups of ionized bands. We compare the spatial distribution of the PCs to the physical conditions, in particular to the strength of the radiation field, $G_{0}$, and the $G_{0}/n_{H}$ ratio and find that the variations in $PC_{2}$, i.e. the ionization state of PAHs are strongly affected by $G_{0}$ whereas the amount of PAH emission (as traced by $PC_{1}$) does not depend on $G_0$.

\end{abstract}

\begin{keywords}
astrochemistry – infrared: ISM – ISM: lines and bands – ISM: molecules
\end{keywords}

\section{Introduction}

Polycyclic Aromatic Hydrocarbons (PAHs) are a large class of complex organic molecules made of carbon and hydrogen. They are strong absorbers of UV photons and release the absorbed energy mainly through vibrational modes in the mid-infrared (MIR) with dominant emission features at 3.3, 6.2, 7.7, 8.6, 11.2, and 12.3 $\mu$m \citep{Sellgren:83, Leger:84, Allamandola:85, Allamandola:89}.

PAHs have been observed in a wide variety of astronomical environments via their characteristic emission features \citep[e.g.][]{Joblin:ngc1333:96, Sloan:99, SmithJD:07, Galliano:08}. To first order, the PAH emission spectrum observed in diverse astronomical environments looks similar. However, there are subtle variations in relative intensities, peak positions, and profile shapes of the emission features depending on the environment in which they are observed \citep[e.g.][]{ Hony:oops:01, Peeters:prof6:02, Galliano:08}. These variations are not only present between the different sources but also spatially within extended sources \citep[e.g.][]{Bregman:05, Peeters:17, Boersma:18}. Various experimental and theoretical studies suggest that the observed variations in the PAH emission features are due to the changes in the properties of the PAH population such as the ionization state, size, and molecular structure \citep[e.g.][]{Allamandola:99, Bauschlicher:08, Bauschlicher:09, Ricca:12, Hony:oops:01, Candian:14}. The most prominent variations are observed in the ratio of 6.2, 7.7, and 8.6 $\mu$m bands to the 11.2 $\mu$m band, which are attributed to the changes in the ionization state of the PAHs \citep[e.g.][]{Allamandola:99, Galliano:08}. Other observed variations include (but are not limited to) variations in the relative intensity of the short to long wavelength PAH bands (e.g. 3.3/11.2) associated with changes in the size distribution of the PAHs \citep[e.g.][]{Schutte:model:93, Ricca:12, Croiset:16, Knight:19}, and variations in the relative intensity of the 11-14 $\mu$m PAH bands associated to the edge structure of the PAHs \citep[e.g.][]{Hony:oops:01, Bauschlicher:08, Bauschlicher:09}.

The observed changes in the characteristics of the PAH population are a result of the changing physical conditions such as density, strength of the UV radiation field, temperature, and metallicity of their residing environments. This clear dependence of PAHs on their local environment makes them a potential tool to probe the physical conditions.
\citet{Galliano:08} and \citet{Boersma:15} used the variations in the ionization state of PAHs to develop a diagnostic tool for tracing the physical conditions. These authors derived an empirical relation between the PAH ionization state (as traced by the 6.2/11.2 band ratio) and the so-called ionisation parameter, $\gamma$ = $G_0 T^{1/2} / n_e$ where $G_{0}$ is the intensity of the radiation field in units of the average interstellar radiation field (the Habing field = $1.6 \times 10^{-3}$ erg cm$^{-2}$ s$^{-1}$), T the gas temperature, and $n_{e}$ the electron density. Recently,  \citet{Pilleri:12} discovered a strong anti-correlation between $G_{0}$ and the fraction of carbon locked up in evaporating small grains (eVSGs). In addition, \citet{Stock:17} reported a relation between $G_{0}$ and the 7.8/7.6 PAH band ratio in Galactic \HII\, regions and reflection nebulae. Although a great deal of work has thus been done to explore the relationships between the PAH variability and the characteristics of their environments, the precise nature of this relationship is not clear yet.

In this work, we use a statistical approach to investigate the variations in the major PAH emission features in the well-known reflection nebula NGC~2023. We perform a Principal Component Analysis (PCA) of a set of fluxes of five major PAH bands at 6.2, 7.7, 8.6, 11.0, and 11.2 $\mu$m to obtain the major driving factors of the observed variations in the PAH emission seen in NGC~2023. The paper is organized as follows. We describe the reflection nebula NGC~2023 in Section~\ref{sec:NGC_2023}. In Section~\ref{sec:PCA}, we briefly summarize the key elements of a PCA. In Section~\ref{sec:results}, we present the results of our PCA of the PAH fluxes, followed by a discussion of PAH sub-populations in Section~\ref{sec:populations} and the peculiar behaviour of the ionic bands in Section \ref{sec:ionic_bands}. Finally, in Section~\ref{sec_physical_conditions}, we discuss our results in the context of the physical conditions in the reflection nebula.

\section{NGC~2023}
\label{sec:NGC_2023}
 NGC~2023 is a bright visual reflection nebula in the Orion constellation at a distance of 403~$\pm$~4 pc \citep{Kounkel:18}. It is illuminated by the young B 1.5V star HD~37903, which carves out a dust-free cavity of $\sim$0.05 pc around itself \citep{Witt:84}, and creates a photodissociation region (PDR) beyond that. For our analysis, we focus on the PAH emission observed in the PDR \citep[e.g.][]{Sellgren:84, Abergel:02, Peeters:12, Shannon:16, Boersma:16, Stock:16,  Peeters:17, Knight:19}.
 
  \begin{figure}
\centering
\includegraphics[scale=0.35]{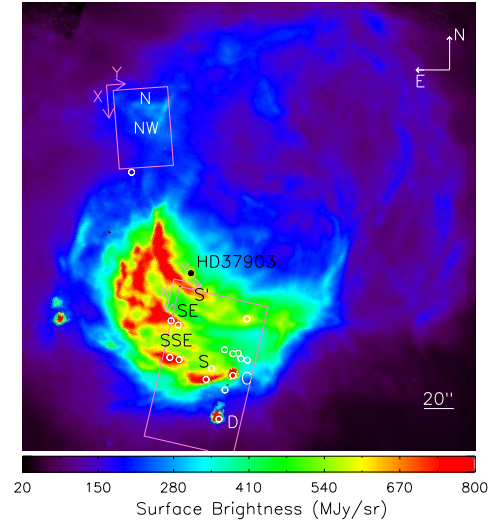} 
\caption{The IRAC [8.0] image of NGC~2023 with the IRS-SL fields of view (FOV) for the north and south regions shown as pink rectangles. The illuminating star HD~37903 is indicated by a black circle. The white circles represent 2MASS point sources, including sources C and D from \citep{Sellgren:83}, located inside the SL FOV. In the south FOV, there are four ridges, the southern ridge (S') at the top of the FOV, the southernmost ridge (S) in the middle of the FOV, the southeastern ridge (SE), and the south-southeastern ridge (SSE). In the north FOV, there are two ridges, the north ridge (N) and the
northwestern ridge (NW). Figure is adapted from \citet{Peeters:17}.}
\label{fig:NGC_2023}
\end{figure}

 We analyzed the MIR data obtained in the two regions of NGC~2023 (see Fig.~\ref{fig:NGC_2023}), hereafter referred to as the north and the south field of view (FOV) respectively, with the Infrared Spectrograph \citep[IRS][]{Houck:04} in the Short-Low (SL) module (Spectral resolution $\sim$ 60-128, pixel scale $\sim$ 1.8$''$), on board the Spitzer Space Telescope \citep{Werner:04a}. The PDR surrounding the illuminating star has bright ridges, which are referred to as the southern ridge (S') at the top of the FOV, the southern ridge (S) in the middle of the FOV, the southeastern ridge (SE), and the south-southeastern ridge (SSE) in the south FOV and the north ridge (N) and the
northwestern ridge (NW) in the north FOV as defined by \citet{Peeters:17}. The north FOV is characterized by a gas density of $\sim10^{4}$ cm$^{-3}$ \citep{Burton:98, Steiman-Cameron:97, Sandell:15} and a radiation field of $\sim10^{3}$ $G_{0}$ \citep{Burton:98, Sandell:15} which are lower as compared to a gas density of $\sim10^{5}$ cm$^{-3}$ \citep{Steiman-Cameron:97, Sheffer:11, Sandell:15} and a radiation field of $\sim10^{4}$ $G_{0}$ \citep{Steiman-Cameron:97, Sheffer:11} in the south FOV.

\section{Principal Component Analysis}
\label{sec:PCA}
A Principal Component Analysis (PCA) is a statistical technique for data visualization and dimensionality reduction developed by \citet{Pearson:1901} and \citet{Hotelling:1933}. It transforms a set of correlated variables in a given data set into a new set of uncorrelated variables called the principal components (PCs) using an orthogonal transformation. The PCs are linear combinations of the original variables obtained  in  such a  way  that  the  first  principal component has the largest variance and contains as much of the statistical information in the data as possible. Each succeeding component has a variance lesser than the preceding component thereby containing the remaining information in the data set. Thus, the goal of a PCA is to find the key variables that can best describe the data. A PCA can be applied to any dataset and is widely used in astronomy. A comprehensive review of PCA can be found in \citet{Abdi:2010} and \citet{Jolliffe:16}.

Here, we briefly describe the underlying mathematical formalism of PCA. Consider a set of $m$ variables \{$x_{i}$\}, measured from a set of $n$ observations in the data set $\textbf{X}$, an $n\times m$ matrix. As a first step, the raw data set, $\textbf{X}$, is standardized such that each variable in the standardized data set, $\mathbf{Z}$,  an $n\times m$ matrix, has a mean of zero and a unit standard deviation. The standardization is done to ensure that all the variables have a comparable scale of measurement so that we get a more meaningful set of PCs. Next, we calculate the covariance matrix, \textbf{C}, an $m\times m$ matrix, of the standardized dataset $\mathbf{Z}$. The covariance is a measure of the degree to which the two variables are correlated. It is zero for two independent variables and is equal to the variance when we calculate the covariance of a variable with itself. The covariance matrix has a set of $m$ eigenvectors, \{$\mathbf{v_{1}}$, $\mathbf{v_{2}}$, ... $\mathbf{v_{m}}$\}, given by the linear combination of the standardized variables:
\begin{equation}
    \mathbf{v_{i}} = a_{i,1}z_{1} + a_{i,2}z_{2} +...+ a_{i,m}z_{m}
\end{equation}

where $a_{i,j}$ are the coefficients of the eigenvector $v_{i}$ and the \{$z_{1}, z_{2}, ...,  z_{m}$\} are the set of original standardized variables. The corresponding set of $m$ eigenvalues, \{$\lambda_{1},$\ $\lambda_{2}, ...$\ $\lambda_{m}$\}, represents the variance of the eigenvectors (for the derivation, see \citet{Jolliffe:16}). Since the key objective of a PCA is to find a set of variables that successively maximizes the variance, the eigenvector corresponding to the largest eigenvalue of \textbf{C} is called the first PC ($PC_{1}$), followed by the eigenvector corresponding to the second largest eigenvalue as the second PC ($PC_{2}$) and so on. Note that the covariance matrix, \textbf{C}, is a symmetric matrix, so the eigenvectors corresponding to different eigenvalues are orthogonal and hence the principal components obtained are independent of each other.
 
The data set in the reference frame of the PCs can be obtained from the following equation: 
\begin{equation}
    \mathbf{Y} = \mathbf{A}\mathbf{Z^{T}}
\end{equation}
where the transpose of the matrix $\mathbf{Y}$ ($\mathbf{Y^{T}}$) is a new data set in the reference frame of the principal components and $\mathbf{A}$ is a transformation matrix given by 
$$
\mathbf{A} =    \begin{bmatrix}
a_{11}&a_{12}&\cdots&a_{1m}\\
a_{21}&a_{22}&\cdots&a_{2m}\\
\vdots & \vdots & \ddots & \vdots \\
a_{m1}&a_{m2}&\cdots&a_{mm}\\

\end{bmatrix}
$$

where the rows of the matrix correspond to the coefficients of the PCs determined from the eigenvalue decomposition of the covariance matrix \textbf{C}, with the first row containing coefficients of $PC_{1}$, the second row containing coefficients of $PC_{2}$ and so on.

\section{PCA of PAH band fluxes in NGC 2023}
\label{sec:results}
\subsection{Measurements of PAH bands}
\label{subsec:Measurements}

To determine the main parameters that drive the variability in the PAH fluxes in NGC~2023, we performed a PCA of the fluxes of five PAH bands at 6.2, 7.7, 8.6, 11.0, and 11.2 $\mu$m in the north and the south FOVs of the nebula. The flux values of these bands are taken from \citet{Peeters:17}. We summarize their measurements here. \citet{Peeters:17} applied three different methods to measure the fluxes of the PAH bands. Here we use the band fluxes measured with the spline decomposition method. In this method, they subtracted a local spline continuum from the spectra and obtained the fluxes of the 6.2, 7.7, and 8.6 $\mu$m PAH bands by integrating the continuum subtracted spectra. The flux measurements of the 11.0 and 11.2 $\mu$m bands, however, were done differently because of their blending with each other. To measure these fluxes, \citet{Peeters:17} fitted these bands with two Gaussians peaking at 10.99 and 11.26 $\mu$m with a FWHM of 0.154 and 0.236 $\mu$m respectively. The Gaussian at 10.99 $\mu$m provides the flux of the 11.0 $\mu$m band. To obtain the flux of the 11.2 $\mu$m band, they subtracted the flux of the 11.0 $\mu$m from the integrated flux of the 11.0 and 11.2 $\mu$m bands in the continuum-subtracted spectra (i.e. not using the Gaussian fit for the 11.2 \mum\, band). Furthermore, the signal-to-noise ratios (SNR) of the PAH bands were estimated using the following expression:
$$
SNR = \frac{I_{PAH}}{rms \times \sqrt{N} \times \Delta \lambda}
$$
where $I_{PAH}$ is the intensity of a PAH band in units of ${\rm W m}^{-2}{\rm sr}^{-1}$, $rms$ is the root-mean-square estimate of the noise, $N$ is the number of spectral wavelength bins in the corresponding PAH band, and $\Delta \lambda$ is the wavelength bin size determined from the spectral resolution. Note that \citet{Peeters:17} take $N$ as the number of data points in the corresponding PAH band. Since the Spitzer IRS data is oversampled by a factor of two, a discrepancy by a factor of $\sqrt{2}$ occurs in the SNR of the PAH band measurements in \citet{Peeters:17} dataset. Various studies show that the plateaus are distinct from the individual PAH bands perched on top of them \citep[e.g.][]{Bregman:89, Roche:orion:89, Peeters:12, Peeters:17}. We have used the measurements of PAH bands from \citet{Peeters:17} that does not include plateaus. Therefore, for any other measurement of PAH bands that does not include plateaus we would expect change in the results smaller than that probed by the dynamic range of PCA. 

The 6.2, 7.7, 8.6, and 11.0 $\mu$m bands are strong in charged PAH molecules whereas the 11.2 $\mu$m band is strong in neutral PAH molecules \citep[e.g.][]{Allamandola:99, Hony:oops:01, Bauschlicher:08}. In our analysis, we have used only the fluxes of the strongest PAH bands except for 11.0 $\mu$m, in order to have good quality measurements. We used the weaker 11.0 $\mu$m band because when normalized to 11.2 $\mu$m, it is a better tracer of the ionization state of PAHs than the other ionized PAH bands \citep[e.g.][]{Rosenberg:11, Peeters:17}. The 11.0 $\mu$m band originates from the out of plane bending modes of solo C-H groups in ionized PAH molecules while the 11.2 $\mu$m  originates from the same mode in neutral PAH molecules, thus the ratio 11.0/11.2 traces solely the ionization state of PAHs without any dependency on other parameters such as the size or the structure of molecule \citep[e.g.][]{Hudgins:99, Hony:oops:01, Bauschlicher:08,Bauschlicher:09}. Moreover, for our analysis, we only included pixels where we have a 3$\sigma$ \footnote[1]{Given the different SNR calculations in \citet{Peeters:17}, our applied 3$\sigma$ limit corresponds to a 4.2$\sigma$ detection.} detection in \textit{all} five PAH bands considered here and masked the remaining pixels. Following \citet{Peeters:17}, we also masked the young stellar objects C and D \citep{Sellgren:83}. Note that we combined the measurements from the north and south FOVs for the PCA analysis discussed in this paper.

\begin{table}
    \centering
    
    \begin{tabular}{c c c}
    \hline
    \multirow{2}{*}{PAH band} & $\langle I_{PAH}\rangle$ & $\sigma_{PAH}$  \\
    & ($\times 10^{-6}$) & ($\times 10^{-6}$)\\
    \hline
    6.2 & 5.705 & 2.550  \\ 
    7.7 & 10.36 & 4.572 \\
    8.6 & 1.694  & 0.833 \\
    11.0 & 0.278  & 0.132 \\
    11.2 & 2.450  & 1.370 \\
    \hline
    \end{tabular}
    \caption{The mean ($\langle I_{PAH}\rangle$) and standard deviation ($\sigma_{PAH}$) values of the PAH band flux variables. All values have units of ${\rm W m}^{-2}{\rm sr}^{-1}$.}
    \label{tab:stats_flux}
\end{table}

We standardized the PAH fluxes of the five bands considered here such that the standardized flux variables have a zero mean and a unit standard deviation

\begin{equation}
    z_{PAH, i} = \frac{I_{PAH, i} - \langle I_{PAH}\rangle}{\sigma_{PAH}}
\end{equation}

where $I_{PAH, i}$ is the intensity of a PAH band at pixel $i$, $\langle I_{PAH}\rangle$ and $\sigma_{PAH}$ are the mean and standard deviation of the measured fluxes of a given PAH band, and $z_{PAH, i}$ is the corresponding standardized intensity. Table \ref{tab:stats_flux} lists the mean $\langle I_{PAH}\rangle$ and standard deviation $\sigma_{PAH}$ values of the PAH flux variables in the nebula. The standardized variables ($z_{PAH}$) have comparable magnitudes and are the input variables in our PCA.

\subsection{Principal Components}
\label{subsec:PCs}

\begin{table}
    \centering
    \begin{tabular}{c c c}
    \hline
    \multirow{1}{*}{PC} & \% variance explained \\
    \hline
    1 & 90.94\\ 
    2 & 7.96\\
    3 & 0.59\\
    4 & 0.44\\
    5 & 0.07\\
    \hline
    \end{tabular}
    \caption{Fraction of variance explained by the principal components (PCs).}
    \label{tab:percent_variance}
\end{table}

Our PCA resulted in five principal components (PCs). These PCs are linear combinations of the standardized flux variables ($z_{PAH}$). We note that the sign of a PC in PCA is arbitrary. Since we will be comparing the PCs (and especially the first two PCs), it makes sense to have them point in the same direction to facilitate interpretation. We thus choose to have the PC$_1$ and $PC_2$ vectors point in the direction of positive $z_{11.2}$ so that larger values of both PC$_1$ and PC$_2$ correspond to larger values of $z_{11.2}$ as well. With that convention, the PCs are determined by the following expressions: 

\begin{equation}
\begin{split} 
PC_{1} = & \,\,  0.466\, z_{6.2} + 0.467\, z_{7.7} + 0.453\, z_{8.6}  \\
& + 0.444\, z_{11.0} + 0.403\, z_{11.2}\\
PC_{2} = & \,\,0.106\, z_{6.2} - 0.016\, z_{7.7} - 0.349 \,z_{8.6}  \\
& - 0.469\, z_{11.0} + 0.804\, z_{11.2}\\
PC_{3} =&\,\, -0.002\, z_{6.2} - 0.288\, z_{7.7} - 0.582 \,z_{8.6}  \\
& + 0.741 \,z_{11.0} + 0.174 \,z_{11.2}\\
PC_{4} =&\,\, 0.461\, z_{6.2} + 0.551\, z_{7.7} - 0.567 \,z_{8.6}  \\
& - 0.141 \,z_{11.0} - 0.378\, z_{11.2}\\
PC_{5} =&\,\, 0.747\, z_{6.2} - 0.629\, z_{7.7} + 0.115 \,z_{8.6}  \\
& - 0.121 \,z_{11.0} - 0.132 \,z_{11.2}
\end{split}
\label{eq:PC_south}
\end{equation}

\noindent
where \{$z_{6.2}$, $z_{7.7}$, $z_{8.6}$, $z_{11.0}$, $z_{11.2}$\} are the standardized flux variables. 

Table \ref{tab:percent_variance} lists the fraction of the variance explained by the individual PCs. Clearly, the first two PCs combined account for the majority of the variation ($\sim$99\%) present in the data. Since the last three PCs combined account for only 1\% of the variation, we exclude those PCs from further analysis. Thus, in the framework of PCA dimensionality reduction, we can now decompose the standardized flux variables into $PC_{1}$ and $PC_{2}$ components:
\begin{equation}
\begin{aligned} 
&z_{6.2} =&  0.466 \,PC_{1} + 0.106 \,PC_{2}  \\
&z_{7.7} =&  0.467 \,PC_{1} - 0.016 \,PC_{2}  \\
&z_{8.6} =&  0.453 \,PC_{1} - 0.349 \,PC_{2}  \\
&z_{11.0} =&  0.444 \,PC_{1} - 0.469 \,PC_{2}   \\
&z_{11.2} =&  0.403 \,PC_{1} + 0.804 \,PC_{2} 
\end{aligned}
\label{eq:z_combined}
\end{equation}

\noindent
These equations include the directional choice discussed above.

\begin{figure}
\centering
\begin{tabular}{c}
 \includegraphics[scale=0.42]{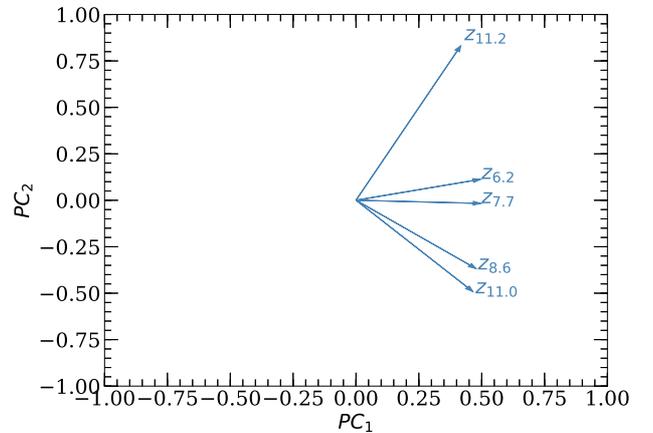}\\
\end{tabular}
\caption{Biplots in $PC_{1}$-$PC_{2}$ plane in NGC~2023. Biplots show the projection of the standardized flux variables onto the $PC_{1}$ and $PC_{2}$ axes.}
\label{fig:biplots}
\end{figure}

Fig.~\ref{fig:biplots} shows so-called ``biplots" obtained from this PCA analysis that show the projection of the standardized original variables ($z_{PAH}$) onto the axes in the new reference frame defined by the PCs. Note that biplots are the visual representation of equation \ref{eq:z_combined}. The larger the projection of a variable on a PC axis, the better that variable correlates with that PC. These biplots are thus useful tools to get a first idea about what drives variations in the original data set. We note that the projection of all $z_{PAH}$ on $PC_{1}$ is in the positive direction indicating that $PC_{1}$ represents changes in total PAH flux. The magnitude of the projection of $PC_{1}$ is roughly the same for all the ionized flux variables ($z_{6.2}$, $z_{7.7}$, $z_{8.6}$, and $z_{11.0}$) but is relatively small for the neutral flux variable ($z_{11.2}$). This means that $PC_{1}$  has a slightly higher contribution from the ionized PAH bands than the neutral PAH band. Note also that $z_{7.7}$ is nearly horizontal in the biplot. Thus, $PC_1$ correlates almost perfectly with $z_{7.7}$, and $z_{7.7}$ is hardly affected by $PC_2$. 

Similarly, the projections of $z_{PAH}$ on $PC_{2}$ provide a hint for the physical interpretation of $PC_{2}$. The projection of $z_{11.2}$ and $z_{6.2}$ on $PC_{2}$ is in positive direction although the projection of $z_{6.2}$ is tiny as compared to that of $z_{11.2}$ while $z_{8.6}$ and $z_{11.0}$ project in the negative direction of $PC_{2}$. The projection of $z_{7.7}$ is almost zero. Due to a clear distinction in the direction of the projection of $z_{8.6}$ and $z_{11.0}$  and $z_{11.2}$, $PC_{2}$ may be interpreted as a tracer of the ionization state of PAHs. However, in this scenario, the positive projection of $z_{6.2}$ is very intriguing and needs further investigation. At the same time, it is worth pointing out that the four ionic bands appear grouped in these biplots: $z_{6.2}$ and $z_{7.7.}$ point in a similar direction, and also $z_{8.6}$ and $z_{11.0}$ point in a similar direction, but the two sets are quite distinct from each other. As will discuss later, this is the first evidence that points to a different character for a subset of the ionic bands.

 \subsection{Spectrum of PCs}
\label{subsec:eigen_spectrum_PCs}

\begin{figure}
\centering
 \includegraphics[scale=0.45]{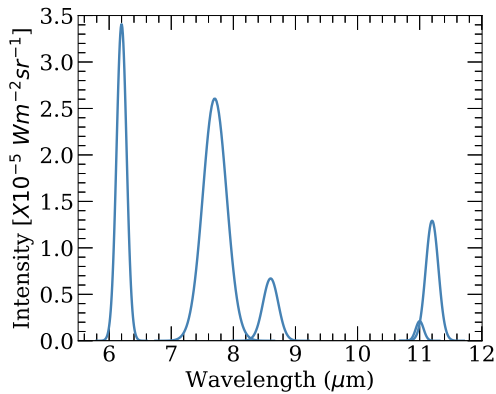}\\
\caption{Characteristic PAH spectrum of $PC_{1}$ in NGC~2023. It is an artificial PAH spectrum created for the five PAH bands considered in the PCA (see text for details). }
\label{fig:PAH_sp_PC1}
\end{figure}

Since $PC_{1}$ is a linear combination of PAH fluxes, the eigen vector corresponding to $PC_{1}$ represents a particular ratio of individual flux values, and thus a characteristic PAH spectrum. In Fig.~\ref{fig:PAH_sp_PC1}, we show the characteristic PAH spectrum of $PC_{1}$. This spectrum is derived from equation \ref{eq:z_combined} by setting $PC_{1}$ = 1 and $PC_{2}$= 0. From the values of the standardized flux variables thus obtained we then extract the actual flux values for the PAH bands by applying an inverse of the standardization operation i.e. adding the mean value of the original flux variables ($\langle I_{PAH}\rangle$) to the product of the standard deviation of the original flux variables ($\sigma_{PAH}$) and the standardized values obtained from equation \ref{eq:z_combined} under the condition of $PC_{1}$ = 1 and $PC_{2}$= 0. The spectrum is then constructed by representing each PAH band by a normalized Gaussian profile at its nominal peak position. Since the width of the observed PAH bands varies from one another, we constructed the Gaussians at 6.2, 7.7, 8.6, 11.0, and 11.2 $\mu$m with a standard deviations of 0.08, 0.19, 0.12, 0.07, 0.10 $\mu$m respectively. The 6.2 and 7.7 $\mu$m bands emerge as strong features. The 8.6 and 11.2 $\mu$m band also have considerable intensities, but significantly lower than those of the 6.2 and 7.7 $\mu$m bands. The 11.0 $\mu$m is the weakest feature because of its weak intrinsic intensity. Theoretical and experimental studies have shown that the spectra of ionized PAH molecules have strong 6.2, 7.7, and 8.6 $\mu$m bands whereas the spectrum of neutral PAH molecules have strong 11.2 $\mu$m band intensity with weak 6.2, 7.7, and 8.6 $\mu$m band intensities \citep[e.g.][]{Allamandola:99, Peeters:prof6:02, Bauschlicher:09}. Thus, the characteristic PAH spectrum for $PC_{1}$ is neither a spectrum typical of solely cations nor solely neutrals but rather that of some mixture of cationic and neutral PAH molecules. The strong contribution from the 6.2, 7.7, and 8.6 $\mu$m bands as compared to the 11.2 $\mu$m in $PC_{1}$ suggest that $PC_{1}$ represents PAH emission of a mixture of PAH molecules where ionized PAHs outweigh the neutral PAHs.

\begin{figure}
\centering
 \includegraphics[scale=0.45]{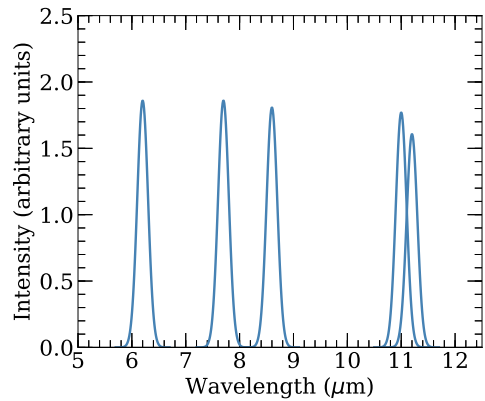}\\
\caption{Eigen spectrum of $PC_{1}$ in NGC~2023. The eigen spectrum is an artificial spectrum created to visualize the contribution of a given PAH band to $PC_{1}$ (see text for details). }
\label{fig:sp_PC1 }
\end{figure}

In order to test our hypothesis and take into account the fact that the 6.2 and 7.7 $\mu$m bands have large intrinsic intensities that can manifest as strong features in the characteristic spectrum of $PC_{1}$, we also constructed a spectrum using only the standardized values of the variables and a standard deviation of 0.1 $\mu$m -- essentially the ``eigen spectrum" corresponding to $PC_{1}$ -- and show it in Fig.~\ref{fig:sp_PC1 }. We note that the eigen spectrum is representative of the eigen vector associated with the PC. In the eigen spectrum of $PC_{1}$, the 6.2, 7.7, 8.6, and 11.0 $\mu$m bands have similar intensities, with the 11.2 $\mu$m band being slightly weaker. This implies that the 6.2, 7.7, 8.6, and 11.0 $\mu$m bands have larger contribution towards $PC_{1}$ as compared to the 11.2 $\mu$m band. Hence, our conclusion about $PC_{1}$ representing PAH emission of a mixture of PAH molecules having more ionized PAHs than the neutrals still holds.

\begin{figure}
\centering
\includegraphics[scale=0.45]{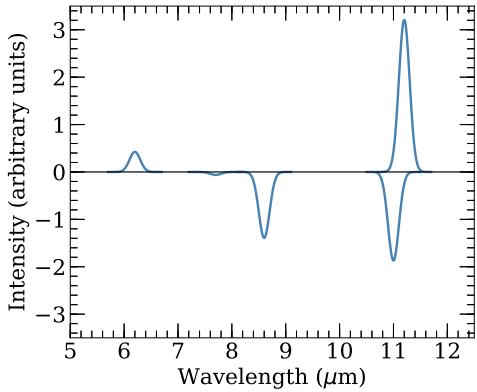}\\
\caption{Eigen spectrum of $PC_{2}$ in NGC~2023. The eigen spectrum is an artificial spectrum created to visualize the contribution of a given PAH band to $PC_{2}$ (see text for details).}
\label{fig:sp_PC2}
\end{figure}
 
We also derived an eigen spectrum of $PC_{2}$. Since $PC_{2}$ represents a first order correction to the PAH fluxes predicted by $PC_{1}$, we only constructed an eigen spectrum of $PC_{2}$ from the standardized flux variables so that we can clearly identify the variations in the relative correction intensities of each of the PAH bands. Setting $PC_{1}$ = 0 and $PC_{2}$ = 1 in equation \ref{eq:z_combined} results in the PAH band intensities corresponding to the eigen spectrum of $PC_{2}$. We emphasize that since the PAH band intensities thus derived are the intensities of the standardized flux variables (with a mean of 0 and a standard deviation of 1), they can have negative values. The resulting eigen spectrum is illustrated in Fig.~\ref{fig:sp_PC2}.

The intensities of the 8.6 and 11.0 $\mu$m  bands are negative, whereas the 6.2 and 11.2 $\mu$m bands have positive intensities. The 7.7 $\mu$m band has almost no intensity. The fact that the 8.6 and the 11.0 $\mu$m bands behave differently than the 11.2 $\mu$m band suggests that $PC_{2}$ is dependent on the PAH ionization state. However, the fact that the 6.2 $\mu$m (strong in ionized PAHs) behave similar as the 11.2 $\mu$m PAH band (strong in neutral PAHs) is very surprising. Indeed, although the 6.2, 7.7, and 8.6 $\mu$m bands correlate very well with each other \citep[see][]{Peeters:17}, their relative contribution to $PC_{2}$ is different with the 6.2 and 7.7 $\mu$m band having opposite contribution with respect to the 8.6 $\mu$m PAH band. This is further addressed in Section~\ref{sec:ionic_bands}.

 \subsection{Correlations between PCs and PAH fluxes}
\label{subsec_correlations_PCs}

\begin{figure*}
\centering
\resizebox{\hsize}{!}{\includegraphics{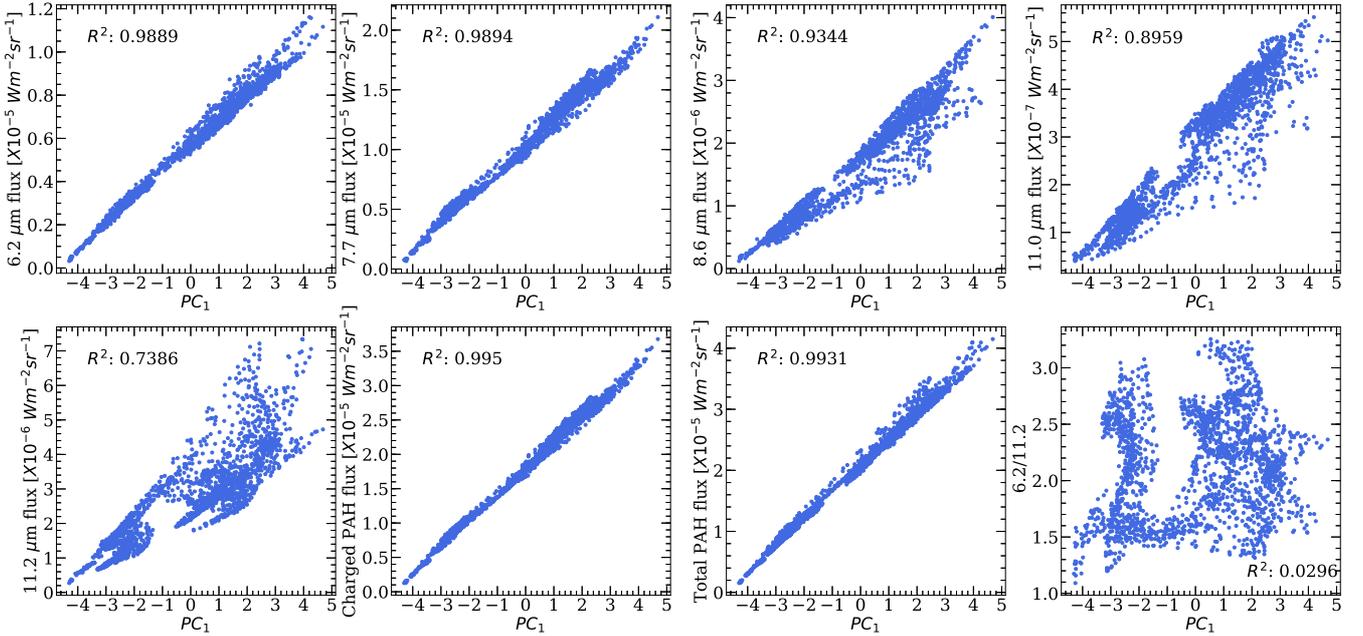}}
\caption{Correlations of $PC_{1}$ with the 6.2, 7.7, 8.6, 11.0, and 11.2 $\mu$m PAH band fluxes, the charged PAH flux, the total PAH flux, and the 6.2/11.2 ratio in NGC~2023. The Pearson correlation coefficient is shown at the top left corner of each plot. $PC_{1}$ is a dimensionless quantity representing the largest variance in the data set.}
\label{fig:correlation_plots_PC1}
\end{figure*}

\begin{table}
    \centering
    \begin{tabular}{c c c }
    \hline
    \multirow{1}{*}{PAH ratio} & R-value & 95\% Confidence Interval  \\
    \hline
    6.2/11.2 & -0.8030 & -0.7866 \, \textemdash  \, -0.8184\\ 
    7.7/11.2 & -0.8217 & -0.8066 \, \textemdash  \, -0.8357\\
    8.6/11.2 & -0.8514 & -0.8386 \, \textemdash  \, -0.8632\\
    11.0/11.2 & -0.8601 & -0.8480 \, \textemdash  \, -0.8714\\
    \hline
    \end{tabular}
    \caption{95\% Confidence Interval of the correlation coefficient (R-value) of $PC_{2}$ with PAH ionization ratios.}
    \label{tab:Confidence_interval}
\end{table}

To gain further insight into the characteristics of the PCs, we investigated the correlations of $PC_{1}$ and $PC_{2}$ with the PAH band fluxes and various PAH band ratios. Fig.~\ref{fig:correlation_plots_PC1} shows the observed correlations for $PC_{1}$ with these variables, and lists their Pearson correlation coefficients. Overall, $PC_{1}$ is well correlated with the individual PAH fluxes and the total PAH flux, albeit with considerable variation in the correlation coefficients. The best correlation is that of $PC_{1}$ with the total charged PAH flux, i.e. the sum of the fluxes of ionized PAH bands (6.2, 7.7, 8.6, and 11.0 $\mu$m) with correlation coefficient of 0.995. This observation is in line with our previous conclusion of $PC_{1}$ tracing emission of a mixture of PAH molecules comprising more ionized PAH molecules than neutrals (we also present the spatial distribution of $PC_{1}$ in Appendix \ref{subsec_spatial_maps_PCs} which reinforces this conclusion). Furthermore, $PC_{1}$ does not correlate at all with any PAH band ratio. As an example of this, we show the correlation of $PC_{1}$ with the 6.2/11.2 band ratio.

We also note some ``branching" in the correlation plots of $PC_{1}$, i.e. there appears to be sets of data points that each correspond to slightly different relationships between parameters. The branching is most prominent in the correlation of $PC_{1}$ with the 11.2 $\mu$m band. Branches are also evident in the correlation of $PC_{1}$ with the 6.2 $\mu$m band and to some extent with the 11.0 and 8.6 $\mu$m bands as well. We will discuss the origin of these branches in Section~\ref{sec:populations}. 

\begin{figure*}
\centering
\resizebox{\hsize}{!}{\includegraphics{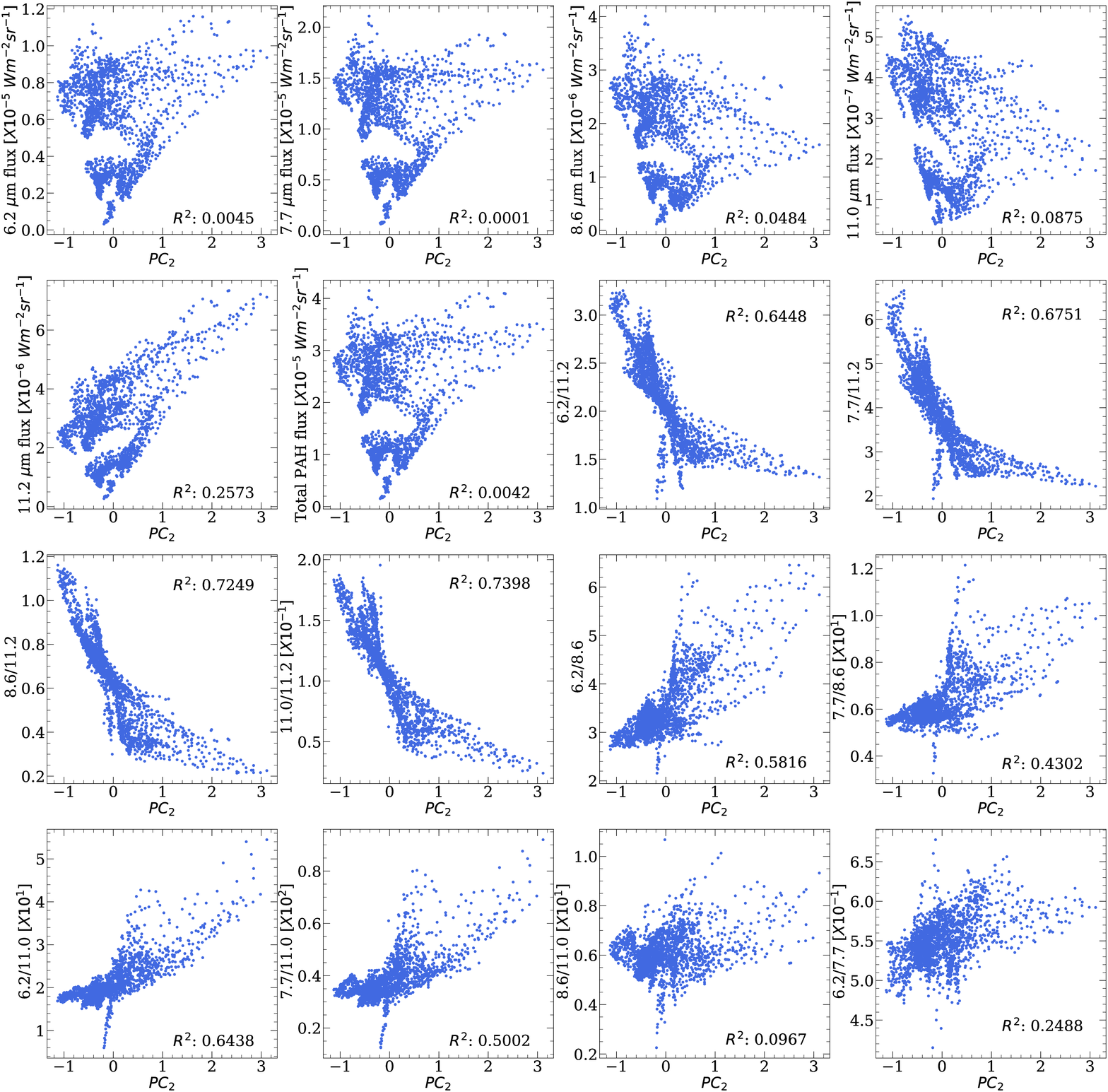}}
\caption{Correlations of $PC_{2}$ with the 6.2, 7.7, 8.6, 11.0, and 11.2 $\mu$m PAH band fluxes, the total PAH flux, and the various PAH band ratios across the south FOV in NGC~2023. The Pearson correlation coefficient is shown in the corner of the each plot. $PC_{2}$ is a dimensionless quantity representing the second largest variance in the data set.}
\label{fig:correlation_plots_PC2}
\end{figure*}

Similar to $PC_{1}$, we also investigated the correlations of $PC_{2}$ with individual band fluxes and various band ratios (Fig.~\ref{fig:correlation_plots_PC2}). $PC_{2}$ does not correlate with individual PAH fluxes nor the total PAH flux. Instead, $PC_{2}$ anti-correlates best with 11.0/11.2, which is a measure of the PAH ionization state. This implies that high $PC_{2}$ values originate from more neutral regions and low $PC_{2}$ values originate from more cationic regions (see also the spatial distribution of $PC_{2}$ in Appendix \ref{subsec_spatial_maps_PCs}). $PC_{2}$ also anti-correlates well with other PAH ratios, the 8.6/11.2, 7.7/11.2, and 6.2/11.2 ratios, but with a decreasing correlation coefficient (ranging from $R^{2}:$ 0.7 to 0.6) in that order. The 11.0, 8.6, 7.7, and 6.2 $\mu$m bands are all attributed to ionized PAHs, so a ratio of any of these PAH fluxes with the 11.2 $\mu$m band traces the ionization state and thus a decrease in the correlation coefficient for these ionic bands requires further investigation. To test the statistical significance of this decrease in the correlation coefficient, we obtained the 95\% confidence intervals (CIs) for each Pearson correlation coefficient (R-value) and checked if there is an overlap between those intervals (Table \ref{tab:Confidence_interval}). The 95\% CI of the correlation coefficient indicates that if we were to repeat our measurements of the PAH bands, we would find that 95\% of the time, the correlation coefficient would fall within this interval. The calculated CIs follow the same trend as the correlation coefficients with a slight overlap between the CIs for 6.2/11.2 and 7.7/11.2 as well as for 8.6/11.2 and 11.0/11.2. We note that the CIs for 6.2/11.2 and 7.7/11.2 do not overlap with the CIs for 8.6/11.2 and 11.0/11.2. Thus the drop in the correlation coefficients of the PAH ionization ratios is not merely an anomaly by chance; rather, it hints towards systematically different behaviour of the ionic bands. We discuss this decrease in the correlation coefficient further in Section~\ref{sec:ionic_bands}. In addition, we find (anti-) correlations of $PC_{2}$ with the other PAH band ratios which are not as tight as those tracing the charge state of PAHs. $PC_{2}$ shows a weak correlation with the 7.7/11.0, 6.2/11.0, 6.2/8.6, and 7.7/8.6 ratios. No (anti-) correlations are found between $PC_{2}$ and the 8.6/11.0 nor the 6.2/7.7 in our dataset.

\section{Evidence for multiple PAH sub-populations}
\label{sec:populations}

\begin{figure*}
\centering
\resizebox{\hsize}{!}{\includegraphics[trim={0cm 0.4cm 0cm 0cm}]{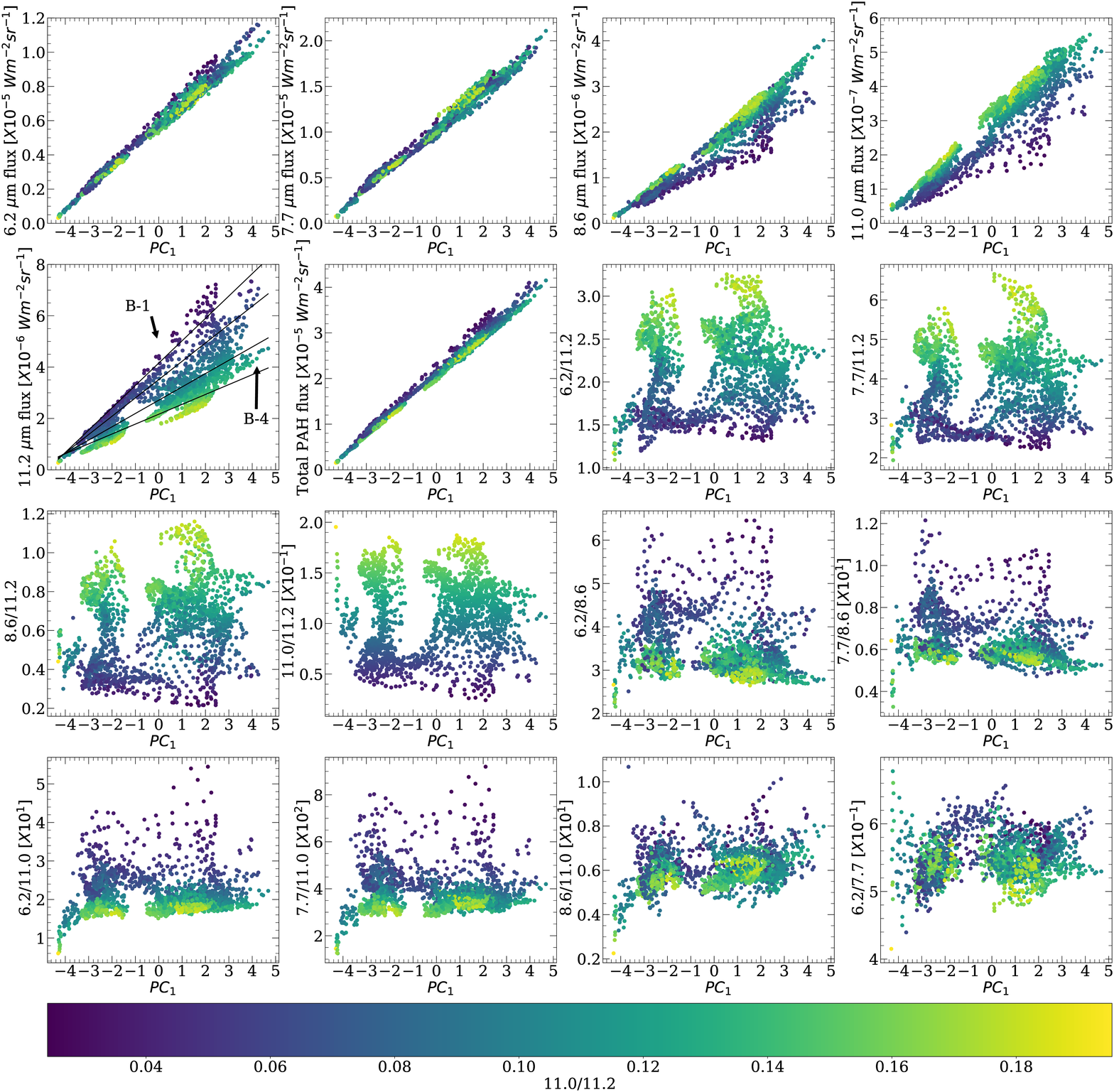}}\\
\caption{Correlations of $PC_{1}$ with the PAH band fluxes and the PAH band ratios color coded based on the 11.0/11.2 ratio to study the origin of the branches, most evident in the $PC_{1}$ - 11.2 correlation (see Section~\ref{subsec_correlations_PCs} for details).}
\label{fig:subpopulations}
\end{figure*}

\begin{figure}
\centering
\includegraphics[scale=0.45]{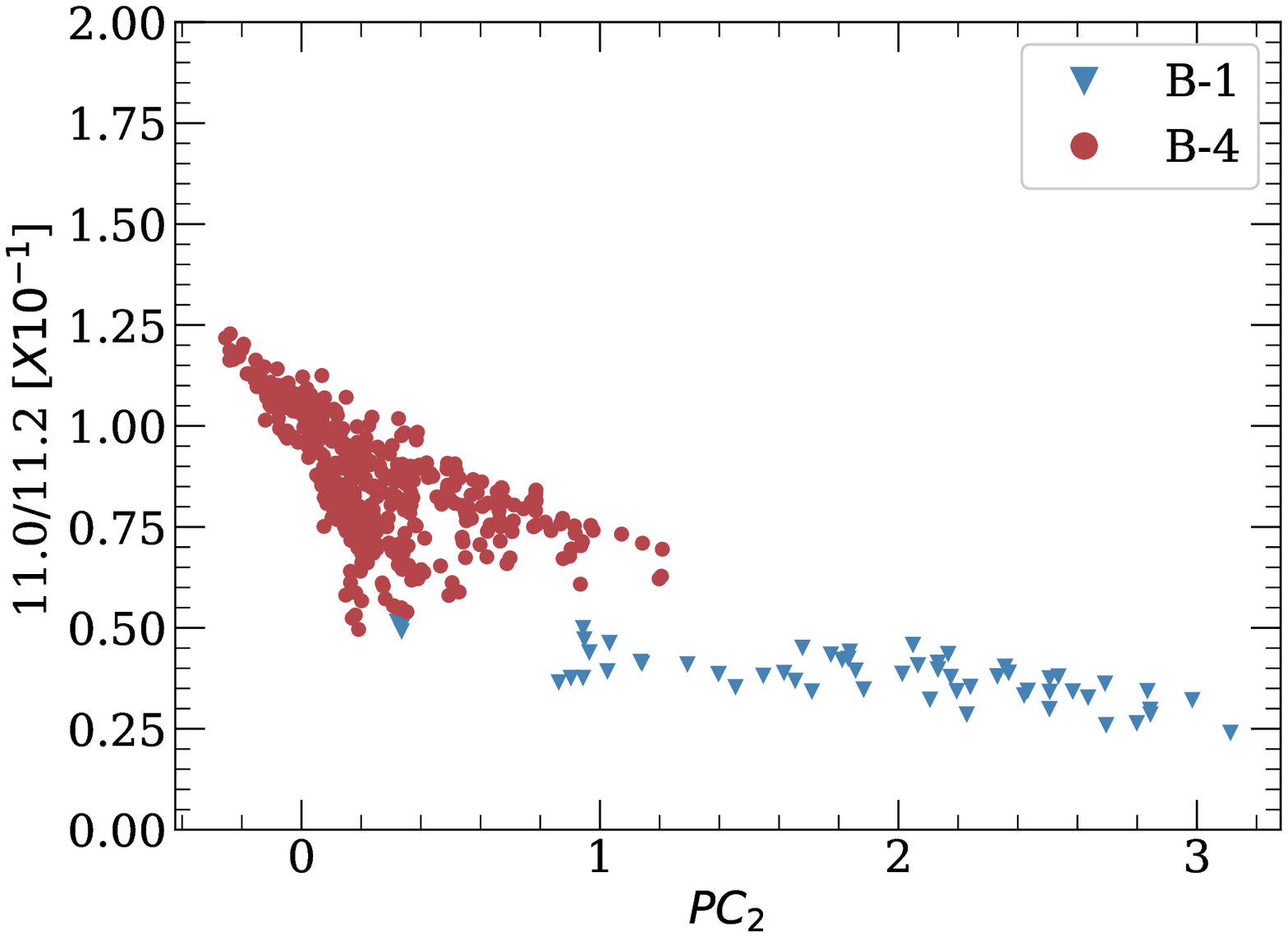}\\
\caption{\textbf{Range of $PC_{2}$ and 11.0/11.2 PAH ratio values within two branches (sub-populations) in NGC~2023.}}
\label{fig:correlation_color_code_PC2 }
\end{figure}

We now return to the branching observed in the correlations of the PCs with the PAH band fluxes (see Section~\ref{subsec_correlations_PCs}). This branching is most prominent in the correlation of $PC_{1}$ with the 11.2 $\mu$m. We show two such branches labelled as $B-1$ and $B-4$ in Fig.~\ref{fig:subpopulations}. A straight line in this correlation represents a set of data points with a specific relationship between the 11.2 $\mu$m PAH flux and the flux of the mixture of PAH molecules traced by $PC_{1}$. The existence of these branches, i.e. a few sets of data points that each are characterized by their own 11.2/PAH flux as traced by $PC_{1}$ ratio, suggests that some of the properties of the underlying PAH populations are very similar within each of the branches, but different across the branches, and hence points towards the presence of distinct PAH sub-populations within the nebula.

In this section, we address the origin of these PAH sub-populations. In order to do so, we color coded the correlations of $PC_{1}$ with all possible PAH band ratios and checked which ratio sets the branches apart. We found that the 11.0/11.2 produces the cleanest separation of the branches (Fig.~\ref{fig:subpopulations}), thereby suggesting that the PAH sub-populations indicated by the branches in the correlation plot of $PC_{1}$ with the 11.2 $\mu$m are the result of the varying PAH ionization across the nebula. We note that while $PC_{2}$ has a strong correlation with the 11.0/11.2 PAH ratio, the branches are not the PAH emission variations traced by $PC_{2}$. Due to the nature of the correlation between $PC_{2}$ and the 11.0/11.2 PAH ratio, $PC_{2}$ values show a relatively wide range of values within each of the branches, although the average value of $PC_{2}$ between the branches is different as well. To illustrate this, we show the dynamic range of $PC_{2}$ and 11.0/11.2 in two of the branches labelled $B-1$ and $B-4$ in Fig.~\ref{fig:correlation_color_code_PC2 }. Note that the Fig.~\ref{fig:correlation_color_code_PC2 } is merely a sub-part of the PC2-11.0/11.2 plot shown in Fig.~\ref{fig:correlation_plots_PC2}. While there are clear differences in ionization (as traced by $PC_{2}$) within a sub-population, their 11.0/11.2 ratio does not change much, and thus the branches or sub-populations are best characterized based on the 11.0/11.2 PAH ratio rather than $PC_{2}$.

There is a gradient in the 11.0/11.2 values across the correlation plots of $PC_{1}$ with the PAH band fluxes. In the $PC_{1}$-11.2 plot, the high 11.0/11.2 values are at the lower part of the envelope of data points and the low 11.0/11.2 values are at the upper part of the envelope of data points as expected. This trend is reversed for the correlations of $PC_{1}$ with the 8.6 and 11.0, i.e. the high 11.0/11.2 values are at the upper part and the low 11.0/11.2 values are at the lower part of the envelope of data points. Surprisingly, this trend is not followed in the $PC_{1}$-6.2 and the 7.7 plot, where there is an increased overlapping of the 11.0/11.2 values.

Since 11.0/11.2 traces the ionization state of PAHs, there is also a gradient of color, as expected, across the y-axis in the correlation plots of $PC_{1}$ with the 6.2/11.2, 7.7/11.2, and 8.6/11.2 PAH band ratios which are also the tracers of the PAH ionization. In addition, there is also an obvious gradient in the 11.0/11.2 values across the y-axis in the $PC_{1}$-6.2/8.6, 7.7/8.6, 6.2/11.0, and 7.7/11.0 plots with high 11.0/11.2 values having low y-values and low 11.0/11.2 values having high y-values. In contrast, we do not observe any gradient in the $PC_{1}$-6.2/7.7 and 8.6/11.0 plots. The presence of the gradient in the 6.2/8.6, 6.2/11.0, 7.7/8.6, and 7.7/11.0 PAH band ratios suggests that there is a distinction between the 6.2, 7.7 $\mu$m bands and the 8.6, 11.0 $\mu$m bands. The absence of any gradient in the 6.2/7.7 and the 8.6/11.0 ratios further suggests that the 6.2 and the 7.7 $\mu$m bands belong to one group of ionic bands and the 8.6 and the 11.0 $\mu$m belong to another. This is further addressed in the next Section.

\section{Peculiar behaviour of the ionic bands}
\label{sec:ionic_bands}
 
By now, we have encountered several instances that suggest that the ionic bands show different behaviour. First, there was the clear separation of the two sets of bands in the biplots. The characteristic spectrum of $PC_{2}$ furthermore shows a very peculiar behaviour of the ionic bands where we found that the behaviour of the 6.2 $\mu$m bears some similarity with that of the 11.2 $\mu$m, a neutral PAH band, while the other ionized PAH bands do not show such similarity. This behaviour is also reflected in the correlation of $PC_{2}$ with the 11.0/11.2, 8.6/11.2, 7.7/11.2, and the 6.2/11.2, where the correlation coefficient decreases in this order, in spite of the 6.2, 7.7, 8.6, and 11.0 $\mu$m bands being attributed to ionized PAHs. Furthermore, the in-depth analysis of the branches seen in the correlations of $PC_{1}$  and $PC_{2}$ reveals that the 6.2 and 7.7 $\mu$m bands behave as one group of ionized bands and the 8.6 and 11.0 $\mu$m bands as another group.

The subtle behaviour of the ionic bands has been recognized previously by several authors \citep[e.g.][]{Galliano:08, Whelan:2013, Stock:14, Peeters:17}. Numerous studies in the literature show that the correlation between the 6.2 and 7.7 $\mu$m bands is stronger than the correlation between the 6.2 or 7.7 and 8.6 $\mu$m bands \cite[e.g.][]{Vermeij:2002, Galliano:08, Peeters:17, Maragkoudakis:2018}. Recently, \citet{Whelan:2013} and \citet{Stock:14} observed the breakdown between the 6.2 and 7.7 $\mu$m bands in two \HII\, regions in the Small Magellanic Cloud and in the Milky Way respectively. Furthermore, the broad 7.7 $\mu$m band is known to have at least two components at 7.6 and 7.8 $\mu$m \cite[e.g.][]{Bregman:89, Cohen:89, Verstraete:2001, Peeters:prof6:02, Bregman:05}. \citet{Rapacioli:2005} argue that the component at 7.8 $\mu$m is due to very small grains (VSGs). More recently, \citet{Bouwman:2019} studied the effect of size, symmetry and structure on the infrared spectra of four PAH cations and noted a drastic change in the vibrational modes of 7-9 $\mu$m region upon the decrease of the molecular symmetry \citep[see also][]{Bauschlicher:09}.

\citet{Peeters:17} did a detailed analysis of emission in the 7-9 $\mu$m region in NGC~2023. They decomposed the spectrum in the 7-9 $\mu$m region into four Gaussian components. These authors found that two of the components centered at 7.6 and 8.6 $\mu$m correlate with each other and with the 11.0 $\mu$m band. These Gaussian components were the main contributors to the traditional 7.7 and 8.6 $\mu$m band intensities. The remaining two Gaussian components centered at 7.8 and 8.2 $\mu$m correlated with each other and displayed a spatial morphology similar to the 11.2 $\mu$m  band and the 5-10 and 10-15 $\mu$m plateau emission in the south FOV and to the 10-15 $\mu$m plateau and the 10.2 $\mu$m continuum emission in the North FOV of NGC~2023. Despite the apparent arbitrariness of the decomposition, their results suggested the presence of at least two distinct sub-populations contributing to the emission in the 7-9 $\mu$m region. In this scenario, the contribution from the Gaussian component at 7.8 $\mu$m to the 7.7 $\mu$m complex is at the origin of the distinction observed between the 7.7 and 8.6 $\mu$m bands detected by the PCA analysis. Thus, the analysis presented here does not explicitly separate the two different PAH populations as suggested by \citet{Peeters:17} in their decomposition of the 7-9 $\mu$m region, rather it provides additional supporting evidence for their existence. Furthermore, our result that the 6.2 and 7.7 \mum\, bands belong to a single group (as opposed to the 8.6/11.0 group), suggests that, similar to the 7.7 \mum\, band, the 6.2 \mum\, band contains contributions of both these two PAH populations responsible for the 7.7 \mum\, band, which \citet{Peeters:17} were unable to extract using their analysis method. This is also supported by the fact that the correlation between the 6.2 and 7.7 \mum\, bands is the strongest \citep[e.g.][]{Peeters:17}.

We further notice that based on the correlations of $PC_{1}$  color-coded with the 11.0/11.2 ratio, the points with low values of 11.0/11.2 populate the regions of high values of 6.2/11.0, 7.7/11.0, 6.2/8.6, and 7.7/8.6. The fact that we see a color distinction in the ratios of these ionic bands due to an ionization ratio (11.0/11.2) itself may suggest that these ratios are further tracing the different ionization states of the PAH molecules. Since high values of 6.2/11.0, 7.7/11.0, 6.2/8.6, and 7.7/8.6 correspond to low values of 11.0/11.2, this then implies that the 6.2 and 7.7 $\mu$m bands originate from less ionized PAHs than the 8.6 and 11.0 $\mu$m bands. Thus, an alternative interpretation for the distinction between the two groups of the ionized bands is that the 6.2 and 7.7 $\mu$m trace singly charged PAH cations and the 8.6 and 11.0 $\mu$m trace doubly charged PAH cations. We emphasize that based on previous studies \citep[e.g.][]{Bauschlicher:09, Peeters:17, Bouwman:2019, Maragkoudakis:2018, Maragkoudakis:2020}, other PAH properties such as size and molecular structure are known to influence the PAH emission spectrum.  \citet{Maragkoudakis:2020} have shown that PAH size primarily effect the 3.3 $\mu$m band and to a lesser extent the 11.2 $\mu$m band indicating that size is likely not the driver of the observed dichotomy between the ionic bands.  The effect of molecular structure on these ionic bands in terms of band assignments has been discussed in \citet{Peeters:17}. However, we currently can not systematically investigate its role within this context based on the astronomical observations.

\section{The PCs and the physical conditions}
\label{sec_physical_conditions}

The primary goal of a PCA is to reduce the set of parameters needed to represent a multivariate data set and find the key variables that drive the input data. By performing a PCA of five PAH band fluxes in NGC~2023, we find that only two variables (PCs) are required to explain $\sim$99\% of the variance in the PAH emission in NGC~2023. Based on the characteristic spectrum of the PCs and their correlations, we conclude that i) $PC_{1}$ representing the largest variance represents the PAH emission of an ion dominated PAH mixture, and ii) $PC_{2}$ constituting the second largest variance has a strong (anti)-correlation with the PAH band ratios, 6.2/11.2, 7.7/11.2, 8.6/11.2, and 11.0/11.2, tracing the ionization state of the PAHs, and a moderate to weak correlation with the 7.7/11.0, 6.2/11.0 and the 6.2/8.6, 7.7/8.6 respectively. In this section, we now explore if there is a relation between the PCs and the parameters that describe the physical conditions in the nebula. Note that the key physical parameters that determine the PAH emission characteristics are the strength of the radiation field, the PAH abundance, electron density, and the temperature. We focus here on the radiation field strength distribution.

We note that the spatial morphology of -$PC_{2}$ resembles that of $G_{0}$ in the north and south FOVs of NGC~2023 (see Fig.~\ref{fig:G0}) with -$PC_{2}$ exhibiting high values in the high $G_{0}$ regions closer to the star (i.e at the bottom of the north FOV and the top of the south FOV) and vice versa. This similarity highlights the influence of $G_{0}$ on $PC_{2}$. Although $PC_{1}$  also exhibits maxima in the high $G_{0}$ regions (S' and SE ridge in south FOV; west of the southern part of the NW ridge), the overall morphology of $PC_{1}$ and $G_{0}$ is very different in both FOVs (see Figs. \ref{fig:spatial_maps_PC1} and \ref{fig:G0} ). Thus we conclude that, variations in the PAH emission reflected by $PC_{2}$ are strongly affected by $G_{0}$, this seems to be less so for $PC_{1}$. We also compared the spatial distribution of PCs to that of $G_{0}/n_{H}$, where $n_{H}$ is the local hydrogen density. \citet{Fleming:2010} presented the map of $G_{0}/n_{H}$ estimated from the ionization state of the PAHs for the south FOV. We find no morphological similarity in the maps of PCs and $G_{0}/n_{H}$ \citep[see Fig. 6 in][]{Fleming:2010}.

In Section~\ref{subsec_correlations_PCs}, we noted that $PC_{2}$ correlated well with the ratios of ionic PAH bands, 6.2/8.6, 7.7/8.6, 6.2/11.0, and 7.7/11.0 (see Figs.~\ref{fig:correlation_plots_PC2}), which reflects the distinction between the two groups of ionic PAH bands. Since $G_{0}$ affects $PC_{2}$ values, one could conclude that the distinction between these two groups of ionic bands is driven by $G_{0}$. This also extends support to our hypothesis that the 8.6 and 11.0 $\mu$m band could be tracing dications as one would expect more doubly charged cations than singly charged cations in high $G_{0}$ regions corresponding to low $PC_{2}$ values and hence low values of 6.2/8.6, 7.7/8.6, 6.2/11.0, and 7.7/11.0 which is indeed the case.

\section{Conclusion}

We have presented a principal component analysis of the fluxes of five major PAH features at 6.2, 7.7, 8.6, 11.0, and 11.2 $\mu$m in the south and the north FOV of NGC~2023. We find that only two principal components (PCs) are required to explain 99\% of the variance in the fluxes of the five PAH bands considered here. Out of the two components, the first PC ($PC_{1}$) is the most important component as it carries the majority (91\%) of the information about the data.

In order to interpret the characteristics of the PCs, we studied their characteristic PAH spectrum, eigen spectrum, and the correlations with the individual PAH band fluxes and PAH band ratios. Based on these we concluded that $PC_{1}$ represents the PAH emission of a mixture of molecules having more ionized PAHs than neutral PAHs and $PC_{2}$ traces the ionization state of PAH molecules. In addition, the correlations of PCs with PAH band fluxes revealed distinct ``branches" which indicated the presence of multiple PAH sub-populations due to varying PAH ionization across the nebula.

Based on the eigen spectrum of $PC_{2}$ and its correlations with the ionic PAH band ratios, we find that there is a distinction between the ionic PAH bands, with the 6.2 and 7.7 $\mu$m bands and the 8.6 and 11.0 $\mu$m bands belonging to two different groups of ionized bands. We further argue that the 6.2 and 7.7 $\mu$m bands originate from less ionized PAHs than the 8.6 and 11.0 $\mu$m bands and thus the 6.2 and 7.7 $\mu$m bands could be attributed to singly charged PAH cations and the 8.6 and 11.0 $\mu$m bands to doubly charged PAH cations. Furthermore, the comparison of PCs with the physical conditions in the nebula shows that the spatial map of -$PC_{2}$ is similar to that of $G_{0}$, and hence we concluded that the $G_{0}$ drives the distinction observed between the ionic bands. 

\section*{Data Availability}
The data underlying this article are available at https://github.com/Ameek-Sidhu/PCA-NGC2023.

\section*{Acknowledgements}
The authors thank the referee for providing valuable comments which led to the improvement of this paper. EP and JC acknowledge support from an NSERC Discovery Grant.

\bibliographystyle{apj}
\bibliography{PAH_papers}

\appendix
\appendixpage
\addappheadtotoc

\begin{appendices}

\section{Spatial Maps of PCs}
\label{subsec_spatial_maps_PCs}

\begin{figure}
\centering

\resizebox{\hsize}{!}{\includegraphics[trim={0cm 0cm 0cm 0cm}, angle=0]{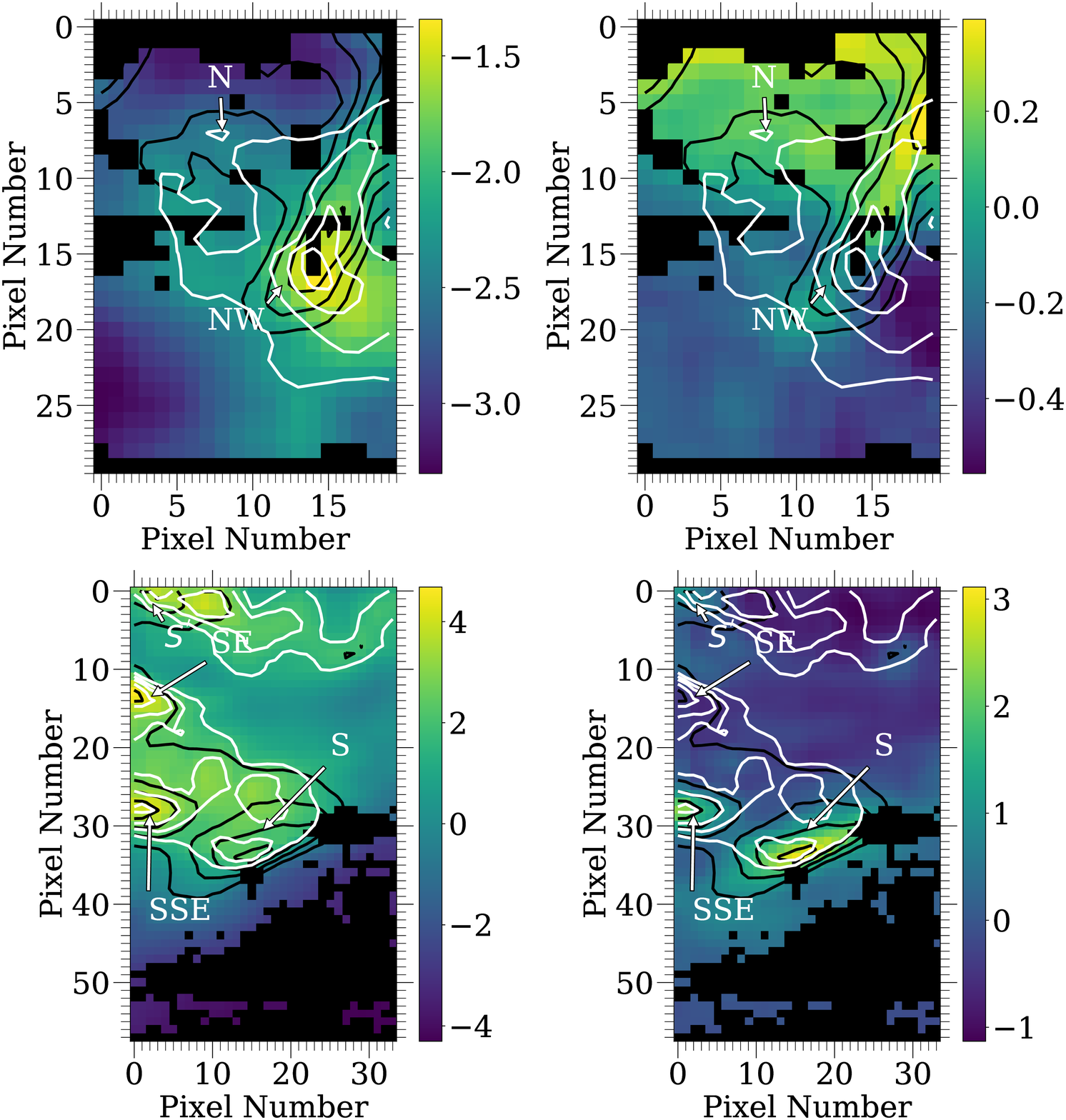}}

\caption{Spatial map of $PC_{1}$ (left) and $PC_{2}$ (right) in the north (top) and the south (bottom) FOV of NGC~2023. For reference, the ridges as defined by \citet{Peeters:17} are annotated and the contours of the intensity of the 7.7 and 11.2 $\mu$m PAH bands are shown in white and black respectively. Pixels without 3$\sigma$ detection for all five PAH bands and containing the YSOs C and D from \citet{Sellgren:83} are masked from the analysis and are shown in black.}.
\label{fig:spatial_maps_PC1}
\end{figure}

It is insightful to create maps of the variations in $PC_{1}$ and $PC_{2}$ and compare those to spatial maps of PAH features described in \citet{Peeters:17}. Fig.~\ref{fig:spatial_maps_PC1} shows the spatial distribution of $PC_{1}$ and $PC_{2}$ in the north and the south FOV. In order to facilitate comparison with the spatial maps of PAH features in \citet{Peeters:17}, we show the contours of the 7.7 and 11.2 $\mu$m PAH bands at the same intensity levels as chosen by \citet{Peeters:17}, i.e. at 1.40, 1.56, 1.70, and 1.90 $\times 10^{-5}$ ${\rm W m}^{-2}{\rm sr}^{-1}$ and 3.66, 4.64, 5.64, and 6.78 $\times 10^{-6}$ ${\rm W m}^{-2}{\rm sr}^{-1}$ respectively. Moreover, we annotate the ridges as defined by \citet{Peeters:17} in the two FOVs.

To first order, the spatial morphology of $PC_{1}$ is very similar to the map of total PAH flux \citep[see Figs. 5 and 6 in][]{Peeters:17}. However, there are subtle differences. In the south FOV, the total PAH emission peaks at S', SE, SSE, and S ridges with the S', SE, and SSE ridges being dominated by cations and the S ridge by neutral PAHs \citep{Peeters:17}. $PC_{1}$ peaks at S', SE, and SSE ridges but not at the S ridge. In the S ridge, $PC_{1}$ values are high, but do not exhibit a maximum. In addition, $PC_{1}$ is strong at the broad, diffuse plateau north and north-west of the S
and SSE ridges, where strong emission is also observed from the 8.6 and 11.0 $\mu$m  PAH emission bands \citep{Peeters:17}. In the north FOV, the total PAH flux peaks on the NW ridge, while $PC_{1}$, although high in the NW ridge, peaks slightly west of the southern part of the NW ridge, where emission from PAH cations peak. Thus, the spatial morphology of $PC_{1}$ in two FOVs shows that $PC_{1}$ peaks in the cation dominated ridges and is high but does not peak in the neutral dominated ridges. This lends further support to our conclusion about $PC_{1}$ that it represents emission of a mixture of PAH molecules with more ionized PAHs than the neutral PAHs.

$PC_{2}$ peaks at the S and the SSE ridge in the south FOV. In all regions other than the S, SSE and a part of the S' ridge, $PC_{2}$ is negative and seemingly uniform. The spatial morphology of $PC_{2}$ resembles that of the H$_2$ 9.7 $\mu$m S(3) and 12.3 $\mu$m S(2) line intensities \citep[see Fig 5 in][]{Peeters:17}. In the north FOV, $PC_{2}$ peaks at the north part of the NW ridge and has a sub-dominant emission in the center part of the N ridge where its projection connects with the NW ridge. $PC_{2}$ exhibits a minimum in the regions south of the N and the NW ridge and west of the southern part of the NW ridge, which are closer to the illuminating star. Its spatial morphology seems to fall between that of the 10-15 \mum\, plateau and the 11.2 \mum\, PAH emission.
The fact that $PC_{2}$ exhibits a maximum in the neutral dominated ridges and a minimum in the cation dominated regions reinforces the suggestion that $PC_{2}$ is a quantity related to the ionization state of the PAHs.

\section{Methodology: Spatial map of radiation field strength}
\label{sec:Methods_Go_maps}
The intensity of the radiation field, $G_{0}$, can be determined indirectly from observations of the FIR continuum emission or the H$_2$ rotation-vibration lines \citep[e.g.][]{Meixner:1992, Black:1987, Draine:1996, Burton:98, Sheffer:11}. It can also be estimated indirectly based on PDR models combined with observations of the FIR cooling lines \citep[e.g.][]{Wolfire:90, Kaufman:99, Sandell:15}. These methods, however, are limited in spatial resolution. Alternatively, we can estimate $G_{0}$ from empirical calibrations where $G_{0}$ is first estimated from other methods and then calibrated against some suitable dust grain or PAH parameter \citep[e.g.][]{Pilleri:12, Stock:17}. Here, we obtained the maps of $G_{0}$ for NGC~2023 using the empirical calibrations of \citet{Stock:17} and \citet{Pilleri:12} (hereafter referred to as method 1 and method 2 respectively) and from observations of the FIR continuum emission (referred as method 3 in the remaining of the paper). These three methods are described in detail below.

\textbf{Method 1}: We derived the morphology of radiation field strength ($G_{0}$) using the empirical relationship established between $G_{0}$ and the ratio of two subcomponents (7.6 and 7.8) of the 7.7 \mum\, feature by \citet{Stock:17}. These subcomponents are two of the four Gaussian components, approximately centered at 7.6, 7.8, 8.2, and 8.6 \mum, used by \citet{Peeters:17} to better fit the features in the 7-9 \mum\, region. Based on a sample of Galactic \HII\, regions and reflection nebulae, \citet{Stock:17} found the following relationship between $G_{0}$ and the 7.8/7.6 ratio:
\begin{equation}
    I_{7.8}/I_{7.6} = (1.70 \pm 0.13) - (0.28 \pm 0.03)\log G_{0}
\label{eq: Stock_method}
\end{equation}

We note that the south FOV of NGC~2023 was in the sample used by \citet{Stock:17} to derive equation \ref{eq: Stock_method}. These authors used slightly different values of central wavelength and full width at half maximum than \citet{Peeters:17} to decompose the features in the 7-9 \mum\, region into the four Gaussian components. Thus, to derive the map of $G_{0}$ in the south FOV we used the fluxes of the 7.6 and 7.8 \mum\ bands obtained from the decomposition parameters of \citet{Stock:17}. The north FOV of NGC~2023 was not part of the sample used by \citet{Stock:17}. We therefore obtained the fluxes of the 7.6 and 7.8 \mum\ bands for this FOV using the average decomposition parameters for the reflection nebulae given by \citet{Stock:17}.

\textbf{Method 2}: The second method used to derive the $G_{0}$ map is based on another empirical relation reported by \citet{Pilleri:12}. These authors found a strong anti-correlation between $G_{0}$ and the fraction of carbon locked in evaporating Very Small Grains ($f_{eVSG}$). eVSGs are carbonaceous grains having a wide size distribution \citep{Li:2001}. It is thought that photo-evaporation of these grains by UV photons lead to the formation of free gas-phase PAH molecules and hence the name eVSGs \citep{Cesarsky:2000, Rapacioli:2005, Berne:2007}. Based on a sample of PDRs, \citet{Pilleri:12} found the following relation:

\begin{equation}
    f_{eVSG} = (-0.23 \pm 0.02)\log G_{0} + (1.21 \pm 0.05)
\label{eq: Pilleri_method}
\end{equation}

which is reliable in the range of $G_{0}$ from $100$ to $5 \times 10^{4}$. 

We obtain $f_{eVSG}$ from the decomposition method PAHTAT, which fits the template spectra of neutral PAHs, ionized PAHs, cluster of PAHs, and eVSGs to the observed spectrum.

\textbf{Method 3}: The third method is based on the FIR continuum emission. This estimation of $G_{0}$ is based on the assumption that all FUV photons are absorbed by dust and re-radiated in the FIR. To measure the FIR flux density, we used the photometric images observed at 70 and 160 \mum\, with the Herschel Photodetector Array Camera and Spectrometer (PACS, \citet{Poglitsch:2010}, AOR Key: ‘PPhoto-ngc2023-135') from the Herschel Science Archive. The PACS 70 and 160 \mum\, filters have pixel scales of 3.2” and 6.4” respectively. In order to achieve the same spatial resolution for both maps, we convolved the 70 \mum\, image to the lower resolution 160 \mum\, image using the convolution kernels and procedures from \citet{Aniano:2011}.  Furthermore, these maps are constrained by the 3$\sigma$ detection limit in the convolved 70 \mum\, and 160 \mum\, images.  

The FIR flux density is then estimated by composing a spectral energy distribution (SED) and then fitting a modified blackbody function to the SED of the form 

\begin{equation}
   I(\lambda, T) = K/\lambda^{\beta} \times B(\lambda, T)
   \label{eq: FIR_flux}
\end{equation}
   
where $K$ is a scaling parameter, $\beta$ is the spectral index, and $B(\lambda, T)$ is the Planck Function as a function of the wavelength ($\lambda$) and the dust temperature ($T$) \citep[e.g.][]{Abergel:2010, Berne:12, Andrews:2018}. To obtain the best fit, we fixed $\beta$ = 1.8 and considered $K$ and $T$ to be the free parameters following previous analysis of similar regions \citep[e.g.][]{Berne:12, Andrews:2018}. The FIR flux is subsequently determined by integrating the area underneath the modified blackbody fit.

Subsequently, we determine $G_{0}$ from this FIR flux measurements following \citet{Meixner:1992}:

\begin{equation}
    G_{0} = 4\pi V l^{-1} S^{-1} \tau \lambda_{0}^{\beta} \int \lambda^{-\beta}  B(\lambda, T) d\lambda
    \label{eq:G0_FIR}
\end{equation}

where $V$ is the volume of the region, $S$ the surface area of the cloud facing the illuminating star, $l$ the pathlength along the line of sight, and $\tau$ the optical depth at the reference wavelength $\lambda_0$. Assuming a spherical geometry for the cavity of NGC~2023 \citep[e.g.][]{Field:1994}, the geometry factor $V l^{-1} S^{-1}$ reduces to 1.0. Thus an estimate for $G_{0}$ is derived by multiplying the FIR flux calculated from equation \ref{eq: FIR_flux} by a factor of 4$\pi$ and converting the units to the Habing field. Note that, the $\tau \lambda_{0}^{\beta}$ term in equation \ref{eq:G0_FIR} is accounted for by the scaling parameter, $K$.\\

\section{Spatial maps of radiation field strength}
\label{sec:Go_maps}

\begin{figure*}
\centering
\resizebox{\hsize}{!}{\includegraphics{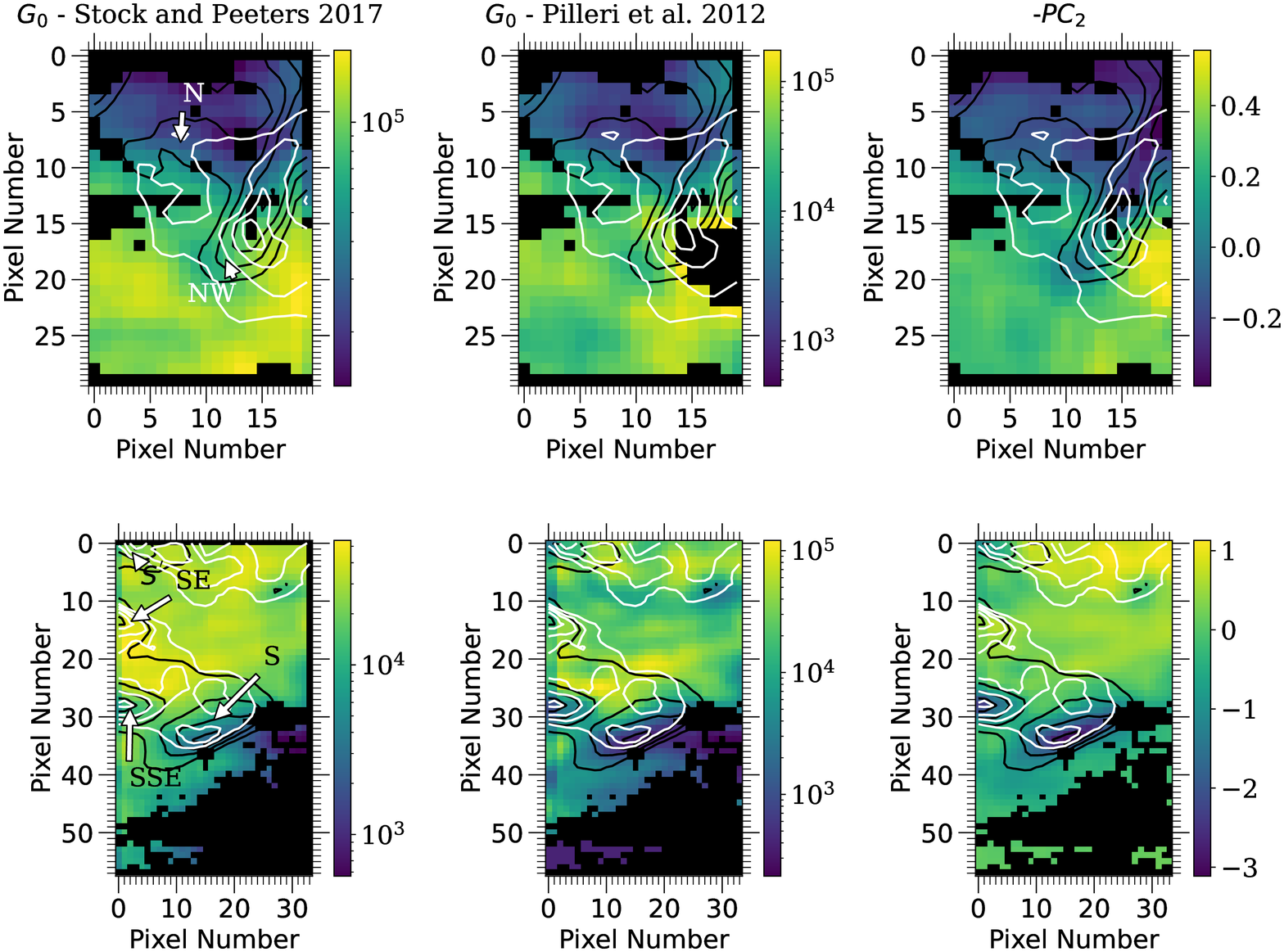}}\\
\caption{Spatial maps of $G_{0}$ in units of Habing Field derived from empirical calibrations across the north (top row) and the south (bottom row) FOV of NGC~2023. Left panel: $G_{0}$ derived from its correlation with the 7.8/7.6 PAH ratio given by
\citet{Stock:17} \textbf{in the logarithmic color scale}. Pixels where the intensity of both 7.6 and 7.8 \mum\ bands is zero are shown in black in addition to those masked while performing PCA; \textbf{Middle panel}: $G_{0}$ derived from its (anti)-correlation with eVSGs given by \citet{Pilleri:12} \textbf{in the logarithmic color scale}. Pixels where $f_{eVSG}$ = 0 are shown in black in addition to those masked while performing PCA: Right panel: $-PC_{2}$ for comparison. Pixels masked while performing PCA are shown in black. For reference, the ridges as defined by \citet{Peeters:17} are annotated for the north and south FOV in left panels. The contours of the intensity of the 7.7 and 11.2 $\mu$m PAH bands are shown in white and black respectively.}
\label{fig:G0}
\end{figure*}

\begin{figure}
\includegraphics[scale=0.4]{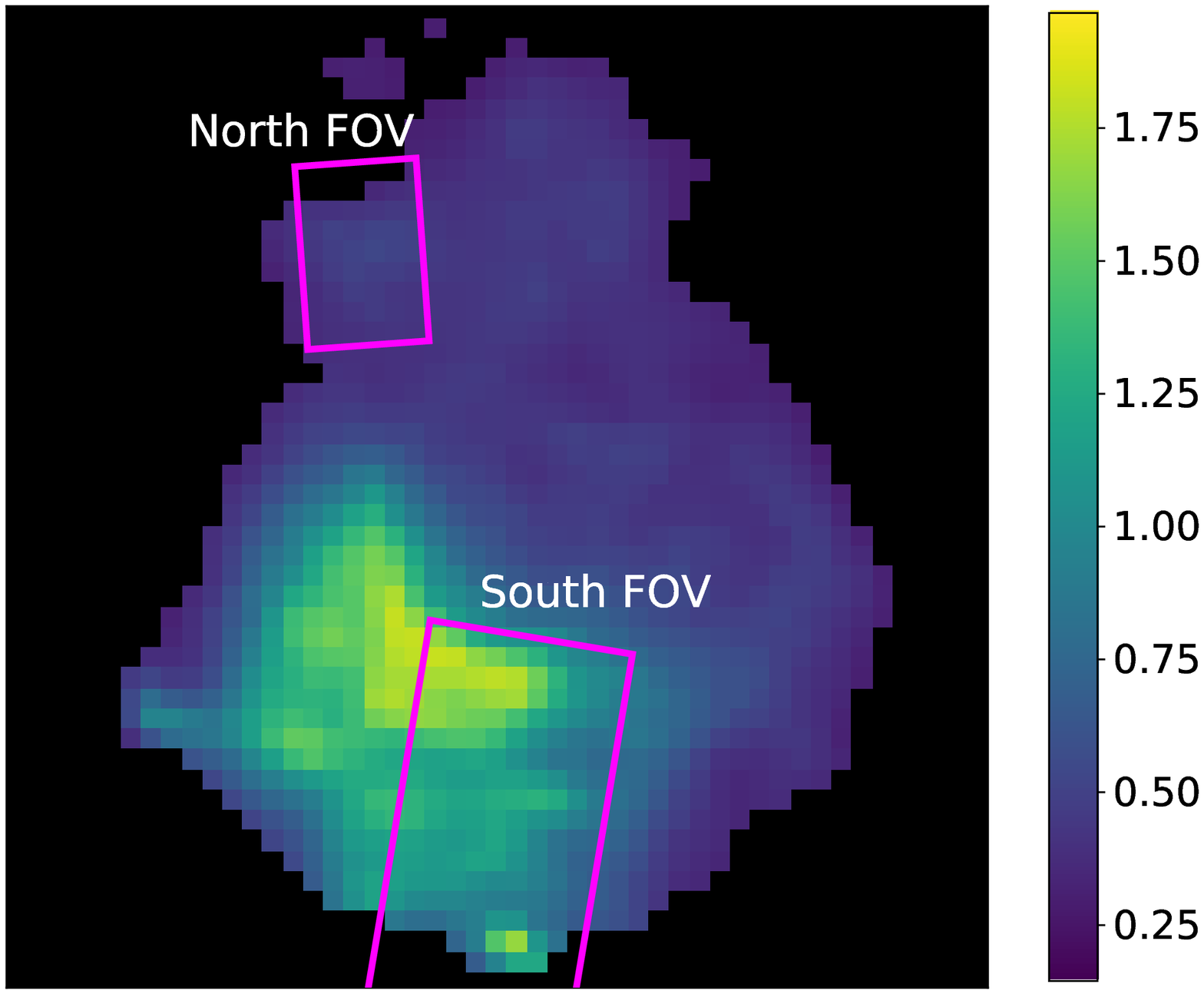}\\
\caption{Spatial maps of $G_{0}$ in units of $10^{4} \times$ Habing Field derived from the FIR continuum measurements (see Appendix \ref{sec:Methods_Go_maps}} for details). North is up and East is to the left in the figure.
\label{fig:G0_FIR}
\end{figure}

Here, we compare the $G_{0}$ maps obtained from the three methods described in Appendix \ref{sec:Methods_Go_maps} with each other in order to have a consistent spatial picture of $G_{0}$ across the nebula. Figs.~\ref{fig:G0} and \ref{fig:G0_FIR} shows the resulting spatial distributions of $G_{0}$ for the north and south FOV based on the three methods. In the south FOV, the absolute values of $G_{0}$ in units of Habing Field estimated from method 1 and method 3 ranges between $10^{3}$ and $10^{4}$, while, those estimated from method 2 varies in the range of $10^{3}$ and $10^{5}$. We emphasize that the $G_{0}$ values $ > 5 \times 10^{4}$ obtained with method 2 are not reliable \citep{Pilleri:12}. The spatial morphology derived from these three methods is fairly similar in the south FOV with high values of $G_{0}$ in the upper half of the FOV containing the S' and the SE ridges and low values in the lower half of the FOV containing the SSE and the S ridges. In the north FOV, the absolute $G_{0}$ values obtained with method 1 and method 2 range from $10^{3}$ to $10^{5}$, whereas those derived from method 3 are of the order of $10^{3}$. We note that the method 1 of estimating $G_{0}$ based on the 7.8/7.6 PAH band ratio is highly sensitive to the decomposition parameters of the 7-9 \mum\ region. Therefore, the high values of $G_{0}$ in the north FOV obtained with method 1 are a consequence of the chosen decomposition parameters. Thus, we conclude that similar to method 2, the high $G_{0}$ values from method 1 are also not reliable in this FOV. Nevertheless, the similar spatial morphology of $G_{0}$ derived from method 1 and method 2 implies that there is a definite variation in $G_{0}$ across the FOV, with $G_{0}$ exhibiting a minimum in the upper part of the FOV containing the N and the northern part of the NW ridge, intermediate values in the south of the NW ridge and a maximum in the lower part of the FOV towards the star. However, due to the limitation in the spatial resolution, a similar conclusion about the spatial morphology could not be drawn for the $G_{0}$ map obtained with method 3. Thus, overall in the south FOV, the three methods of estimating $G_{0}$ are consistent with each other in terms of spatial distribution and the absolute values except for the high values of $G_{0}$ in method 2. On the other hand in the north FOV, the absolute $G_{0}$ values derived from these three methods are not comparable. While the information on spatial morphology of $G_{0}$ from method 3 could not be obtained, we find that method 1 and method 2 are consistent with each other in this FOV.

\end{appendices}

\bsp	
\label{lastpage}
\end{document}